\documentclass[10pt,letterpaper]{article}
\usepackage[top=0.85in,left=1.5in,footskip=0.75in,marginparwidth=1in]{geometry}

\usepackage[utf8]{inputenc}

\usepackage{cite}

\usepackage{nameref,hyperref,cleveref}

\usepackage[right]{lineno}

\usepackage{microtype}
\DisableLigatures[f]{encoding = *, family = * }

\raggedright
\usepackage{changepage}

\usepackage[aboveskip=1pt,labelfont=bf,labelsep=period,singlelinecheck=off]{caption}

\makeatletter
\renewcommand{\@biblabel}[1]{\quad#1.}
\makeatother

\usepackage{lastpage,graphicx}
\usepackage{epstopdf}
\usepackage{color}

\definecolor{Gray}{gray}{.25}

\usepackage{graphicx}

\usepackage{sidecap}

\usepackage{wrapfig}
\usepackage[pscoord]{eso-pic}
\usepackage[fulladjust]{marginnote}
\reversemarginpar

\begin{document}
\vspace*{0.35in}

\begin{flushleft}
{\Large
\textbf\newline{ Higher flow harmonics of strange hadrons in Au+Au collisions at $\sqrt{s_{NN}}$ = 200 GeV and Pb+Pb collisions at $\sqrt{s_{NN}}$ = 2.76 TeV with HYDJET++ model}
}
\newline
\\
Gauri Devi\textsuperscript{1},
Arpit Singh\textsuperscript{1},
Saraswati Pandey\textsuperscript{1},
B. K. Singh\textsuperscript{1,2*},

\bigskip
\bf{$^{1}$}Department of Physics, Institute of Science, Banaras Hindu University (BHU), Varanasi, 221005, India 
\\
\bf{$^{2}$}Discipline of Natural Sciences, PDPM Indian Institute of Information Technology Design \& Manufacturing, Jabalpur, 482005, India
\\
\bigskip
gauri.devi13@bhu.ac.in \\
arpit.singh@bhu.ac.in \\
saraswati.pandey13@bhu.ac.in \\
$*$bksingh@bhu.ac.in

\end{flushleft}

\begin{abstract}
\noindent
Using the HYDJET++ model, we measure the higher-order flow harmonics $v_{n}$(n=2,3,4) of (multi-) strange hadrons in Au+Au collisions at $\sqrt{s_{NN}}$ = 200 GeV and Pb+Pb collisions at $\sqrt{s_{NN}}$ = 2.76 TeV. We have compared our model results and experimental data at RHIC and LHC energies. The model reproduces the higher flow harmonics of (multi-) strange hadrons as a function of $p_{T}$, centrality, and quark content. We have studied and discussed mass ordering of flows $v_{n}$(n=2,3,4) among $\pi^{+}+\pi^{-}$, $K^{-}+K^{+}$, $K_{s}^{0}$, $p+\overline{p}$, $\Lambda+\overline{\Lambda}$, $\Xi^{-}+\overline{\Xi}^{+}$, $\Omega^{-}+\overline{\Omega}^{+}$ at low $p_{T}$ and the baryon-meson grouping at intermediate $p_{T}$. NCQ scaling using the HYDJET++ model and its comparison with experimental results at RHIC and LHC energies have also been discussed. Study of flow harmonics behaviour in RHIC and LHC energy regimes will provide an additional insight into the dynamics of anisotropic flow and the effect of radial flow expansion in the system.
\end{abstract}

\section{Introduction}\label{Introduction}
The relativistic heavy-ion collision experiments aim to explore the Quark-Gluon Plasma (QGP) formed at high densities and temperatures \cite{ALICE:2021ibz, Snellings:2011sz, Snellings:2014vqa, STAR:2021twy, STAR:2005gfr, Voloshin:2008dg, Voloshin:1994mz}. Among various signals~\cite{Singh:1992sp} suggested to probe the QGP matter, anisotropic flow~\cite{Stoecker:2004qu} serves as a powerful tool in non-central heavy-ion collisions to study the properties and dynamics of QGP. The anisotropic flow arises in the final hadronic distribution from the spatial asymmetry in the initial geometry of the collision transformed into the momentum anisotropy combined with the initial state inhomogeneities of the system's energy density. The anisotropic flow is calculated in the form of Fourier series expansion with respect to the event plane as:

\begin{equation}
\frac{dN}{d\phi} = 1 + 2 \sum_{n=1}^{\infty}  v_{n} \cos[n(\phi-\psi_{n})]
\end{equation}

The $n^{th}$ order flow harmonics $v_{n}$ are represented by the Fourier coefficients as :
\begin{equation}
v_{n} =  \langle  \langle \cos [n(\phi-\psi_{n})] \rangle \rangle
\end{equation}

where $\phi$ is the azimuthal angle of the produced particles, n is the harmonic order of flow coefficients and $\psi_{n}$ is the azimuthal angle of the reaction plane. The averaging is performed over all the particles in a single event and over all the events. The elliptic flow $v_{2}$ arises from the almond-shaped geometry of the interaction volume being, the largest contribution to the asymmetry of non-central collisions. The triangular flow $v_{3}$ arises from fluctuations in the initial distribution of particles \cite{STAR:2013qio, Solanki:2012ne, ALICE:2016cti, STAR:2010ico}. The quadrangular flow $v_{4}$ arises from both, the asymmetry of non-central collisions and from fluctuations. The higher flow harmonics $v_{n}(n> 2)$ arise from the event-by-event fluctuations in the initial distribution of nucleons in overlap region and non-viscous effects \cite{Heinz:2013bua}, \cite{Bravina:2015sda}. They are more sensitive to transport coefficients ($\eta/s$) than elliptic flow \cite{STAR:2021twy, ALICE:2016cti}. 

In the past few decades, different facets of (multi-) strange hadrons have been explored theoretically \cite{Tiwari:1997zu,Bazavov:2014xya,STAR:2005gfr,PhysRevC.107.024906} and experimentally \cite{ALICE:2021ibz,STAR:2021twy,STAR:2015gge,STAR:2022ncy,ALICE:2014wao,STAR:2005gfr,STAR:2005npq,STAR:2008ftz,STAR:2022tfp} in heavy-ion collisions. The main motivation of our presented work is to study anisotropic flow $v_{n}$(n=2,3,4) of (multi-) strange hadrons. Strange hadrons, especially having more than one valence strange quark, are considered as a good probe to study the collective properties of the medium created in high energy heavy-ion collisions \cite{STAR:2022tfp, PhysRevC.107.024906}. Anisotropic flow harmonics are transverse momentum ($p_{T}$) dependent and produce a hump structure \cite{ALICE:2018rtz, Bravina:2013xla, Lokhtin:2012re}. It increases monotonically with $p_{T}$ up to 2-3 GeV/c where a maximum is achieved and then decreases. The position of this maxima depends on collision centrality and particle species \cite{ALICE:2021ibz, ALICE:2014wao, ALICE:2018rtz}. Flow harmonics also show mass ordering at low $p_{T}$ i.e., $v_{n}^{mesons} > v_{n}^{baryons}$~\cite{ALICE:2018yph,ALICE:2016ccg}. A characteristic mass dependence of $p_{T}$- differential $v_{n}$ occurs as a consequence of the interplay between the azimuthally symmetric radial flow  and anisotropic flow, $v_{n}$ (n$\geq$ 2). Mass ordering in anisotropic flow at  RHIC (the Relativistic Heavy Ion Collider) \cite{STAR:2004jwm, PHENIX:2014uik}, and LHC (the Large Hadron Collider)  \cite{ALICE:2018lao, ALICE:2011ab} has been observed at $p_{T}<$ 2-3 GeV/c \cite{ALICE:2014wao, ALICE:2021ibz, Zhu:2016qiv}. At high $p_{T}$, particle-type dependence was observed i.e., $v_{n}^{baryons}>v_{n}^{mesons}$~\cite{STAR:2005gfr,STAR:2022ncy,ALICE:2014wao,ALICE:2021ibz,STAR:2021twy,Pandey:2021ofb}. 

The collective behaviour of (multi-) strange hadrons is weaker than that of the light strange and non-strange hadrons. Since the hadronic interaction cross-section of multi-strange hadrons with their large mass is expected to be small, their freeze-out temperatures are close to the quark-hadron transition temperature as predicted by lattice QCD \cite{STAR:2021twy, STAR:2022tfp, STAR:2015gge}. Other studies on the behaviour of strange hadrons based on temperature effects have been performed using the hydrodynamic-inspired models~\cite{Schenke:2019ruo}. It shows that (multi-) strange baryons thermally freeze-out close to the point where chemical freeze-out occurs with $T_{ch}$=160-165 MeV \cite{STAR:2005npq}, which at these collision energies coincides with the critical temperature $T_{c}$. In addition, according to the Quark-Coalesence mechanism \cite{Voloshin:2002wa}, \cite{Molnar:2003ff}, strange quarks may have a smaller flow than light quarks at high $p_{T}$ because heavy strange quarks are expected to lose less energy in nuclear medium \cite{Molnar:2003ff}. Thus, the flow harmonics for strange and non-strange hadrons have same qualitative behaviour for u, d, and s quarks; however, the strength could be comparable~\cite{Bazavov:2014xya, STAR:2022ncy}. Moreover, compared to non-strange hadrons, flow for (multi-) strange hadrons provide more information about the partonic stage than the hadronic stage due to the small hadronic cross-section of strange hadrons. This can be visualized by studying NCQ scaling~\cite{STAR:2015gge, STAR:2022ncy, ALICE:2014wao} of anisotropic flow. 

In recent times, studies have been performed on the flow coefficients of all charged (both strange and non-strange) and identified hadrons at RHIC~\cite{STAR:2021twy, STAR:2015gge, STAR:2022tfp, STAR:2022ncy, Singh:2017fgm} and  LHC~\cite{ALICE:2021ibz, ALICE:2014wao, ALICE:2018rtz, ALICE:2016ccg, Pandey:2021ofb, Schenke:2020mbo, ALICE:2018lao} using different models such as AMPT \cite{STAR:2021twy, STAR:2015gge}, \cite{Solanki:2012ne}, VISHNU \cite{Zhu:2016qiv, Zhu:2015dfa, Bass:1998ca}, ideal Hydrodynamic model \cite{STAR:2021twy, Qiu:2011hf}, and HYDJET++ \cite{Bravina:2013xla, Singh:2017fgm, PhysRevC.103.014903, Pandey:2021ofb, Bravina:2020sbz,Zabrodin:2016wmo, Crkovska:2016flo} models using by several analysis methods like reaction-plane method $v_{n}\{RP\}$, sub-event plane method $v_{n}^{sub}\{EP\}$, multi-particle cumulant method $v_{n}\{2,3,4\}$, and LYZ method $v_{n}\{LYZ\}$ \cite{Bazavov:2014xya,Solanki:2012ne, Zhu:2016qiv, Zhu:2015dfa, CMS:2013wjq, ALICE:2016cti, Voloshin:1994mz}. 

In this article, we have used the Monte Carlo HYDJET++ model to analyze higher flow harmonics ($v_{2}-v_{4}$) as a function of transverse momentum and centrality for (multi-) strange hadrons at RHIC and LHC energies. A brief description of HYDJET++ model is discussed in section \ref{model}. In section \ref{Results}, results obtained by HYDJET++ model on anisotropic flow coefficients are discussed along with its mass ordering and NCQ scaling ($n_{q}$ is the number of constituent quarks). At last, we have presented the summary of our study in Section \ref{conclusion}.
\section{Model}\label{model}
HYDJET++ model is a Monte Carlo event generator that simulates relativistic heavy-ion collisions in an event-by-event basis \cite{Lokhtin:2005px, Lokhtin:2009hs, Lokhtin:2010zz}. A detailed description of the used physics and the simulation procedure is provided in the HYDJET++ manual \cite{Lokhtin:2008xi, Lokhtin:2009hs}. HYDJET++ is primarily composed of two independent components which simulate soft (low $p_{T}$ physics) and hard processes (high $p_{T}$ physics) \cite{Lokhtin:2010zz}. The soft part provides the hydrodynamical evolution of the system while the hard part describes multiparton fragmentation within the formed medium. The soft state in HYDJET++ is a thermal hadronic state generated on the chemical and thermal freezeout hypersurfaces obtained from a parameterization of relativistic hydrodynamics with preset freezeout conditions provided by the FASTMC~\cite{Amelin:2006qe, Amelin:2007ic} generator. The soft part includes the generation of the 4-momentum of hadrons, fluid flow 4-velocity, and the two- and three-body decays of resonances with branching ratios taken from the SHARE~\cite{Torrieri:2004zz} particle decay table. The hard part of the model consists of PYTHIA \cite{Sjostrand:2006za} and PYQUEN \cite{Lokhtin:2005px} event generators. These generators simulate initial parton-parton collisions, the radiative energy loss of partons, and parton hadronization. The model has several free parameters for the soft component and can be fixed by fitting multi-parameters at the preset freezeout surface and calculating various observations at RHIC and LHC.

\begin{figure*}[htbp]
\includegraphics[width=0.92\textwidth] {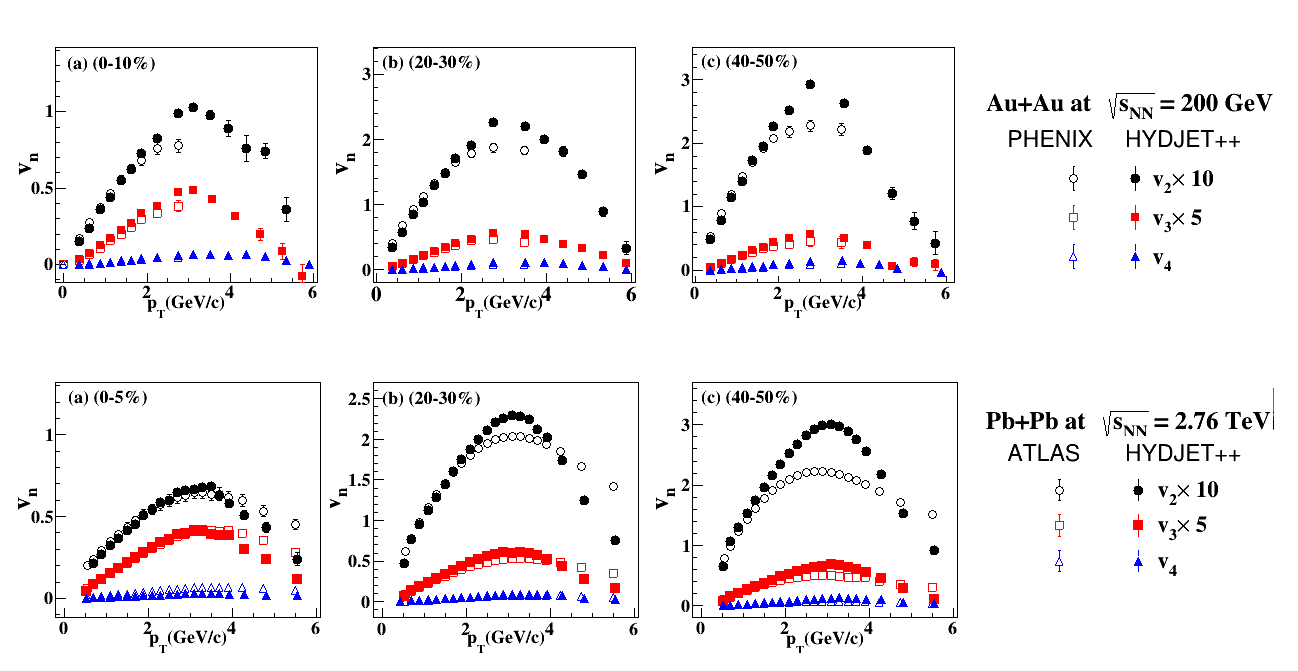}
\vspace{-1mm}
\caption{ The first row ((a)-(c)) represent the higher harmonic coefficients $v_{n}$ of charged hadrons at mid-rapidity ($|y|<$ 1.0) as a function of transverse momentum $p_{T}$ at $\sqrt{s_{NN}}$= 200 GeV in Au+Au collisions and the second row ((a)-(c)) represent $v_{n}$ of charged hadrons at mid-rapidity ($|y|<$ 0.5) with respect to $p_{T}$ at $\sqrt{s_{NN}}$= 2.76 TeV in Pb+Pb collisions. The solid markers represent the HYDJET++ model results while the open markers represent experimental data from the PHENIX~\cite{PHENIX:2011yyh} and ATLAS~\cite{ATLAS:2012at} collaborations.}
\label{fig1}
\end{figure*}

The hard part includes the generation of multi-jets according to the binomial distribution in PYQUEN. PYQUEN includes generating initial parton spectra with PYTHIA and production vertex at a given impact parameter followed by the rescattering-by-rescattering simulation of the parton path in a dense zone where the partons experience radiative and collisional energy losses. Then, the final hadronization of hard partons and in-medium emitted gluons takes place in  accordance with the Lund string model. The average number of jets produced in an AA event collision is calculated  as a product of the number of binary NN collisions at a given impact parameter per the integral cross-section of the hard process in NN collisions with the minimum transverse momentum transfer $p_{T}^{min}$. 

Within the soft part of HYDJET++, anisotropic flow arises from the corresponding spatial eccentricities of the overlap zone of the fireball. The flow fluctuations in HYDJET++ arise from fluctuations in particle multiplicity, decays of resonances, and the production of (mini-)jets \cite{Bravina:2013xla},\cite{Bravina:2015sda}. The second harmonic coefficient $v_{2}$ is calculated using the initial state coordinate anisotropy($\epsilon_{0}(b)$) parameters and momentum anisotropy ($\delta(b)$) parameters in the model. The third harmonic coefficient $v_{3}$ is calculated by the spatial triangularity parameter $\epsilon_{3}(b)$. These parameters can be treated either independently for each centrality or related to each other through the dependency on initial ellipticity $\epsilon_{0}(b) = b/2R_{A}$, where $R_{A}$ is the nucleus radius. During the evolution, the initial system coordinates anisotropy $\epsilon_{0}(b)$ is converted into the momentum anisotropy $\delta(b)$ in the transverse plane at the freezeout surface \cite{Lokhtin:2010zz}. The anisotropy eccentricity parameters are related to the transverse radius of the fireball by the equations :

\begin{equation}
R_{ell}(b,\phi)= R_{eff}(b) \frac{\sqrt{1-\epsilon^2(b)}}{1+ \epsilon(b)\cos(2\phi)} 
\end{equation}
and,
\begin{equation}
R_{trian}(b,\phi) = R_{ell}(b,\phi)[1+\epsilon_{3}(b)\cos[3(\phi-\psi_{3})]]
\end{equation}

respectively. Experiments have shown that there is no correlation between the event planes of the second and third harmonics. All produced particles are emitted from a thermal expanding source with a maximum transverse flow velocity $\rho_{u}$ whose parametrization is presented in reference~\cite{Lokhtin:2008xi}.
\begin{equation}
 \rho_{u}^{max} = \rho_{u}^{max}(0) [1+ \rho_{2u}(b)\cos(2\phi)+ \rho_{3u}(b)\cos(3(\phi-\psi_{3})
 + \rho_{4u}(b)\cos(4\phi)]
\end{equation}

This also permits us to find the higher order anisotropic flow coefficients with respect to the reaction plane i.e., $v_{2,4} (\psi_{2}^{RP} =0)$ and  $v_{3}(\psi_{3}^{RP})$. Here, we perform the modulation of the transverse velocity profile in all freeze-out hypersurfaces. The new parameters $\rho_{3u}(b)$ and $\rho_{4u}(b)$ can also be treated either centrality independent or dependent on initial ellipticity. We have used the former way for our simulation and chosen a fixed value of these parameters  for Au+Au collisions at $\sqrt{s_{NN}}$ = 200 GeV and Pb+Pb collisions at $\sqrt{s_{NN}}$ = 2.76 TeV, respectively \cite{Bravina:2013xla}.

\begin{figure*}[htbp]
\includegraphics[width=0.815\textwidth]{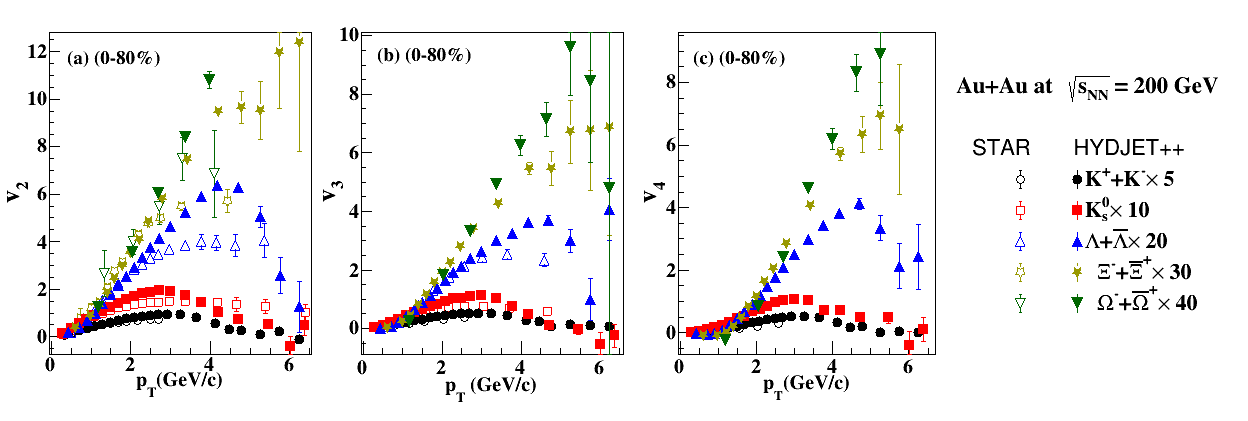}
\caption{The transverse momentum $p_{T}$ dependence of $v_{2}$ (a), $v_{3}$ (b) and $v_{4}$ (c) of the (multi-) strange hadrons for  $(0-80)\%$ centrality in Au+Au collisions at $\sqrt{s_{NN}}$= 200 GeV. The solid markers represent the HYDJET++ model results while the open markers represent experimental data from the STAR collaboration~\cite{STAR:2022ncy, STAR:2008ftz}.}
\label{fig2}
\includegraphics[width=0.805\textwidth] {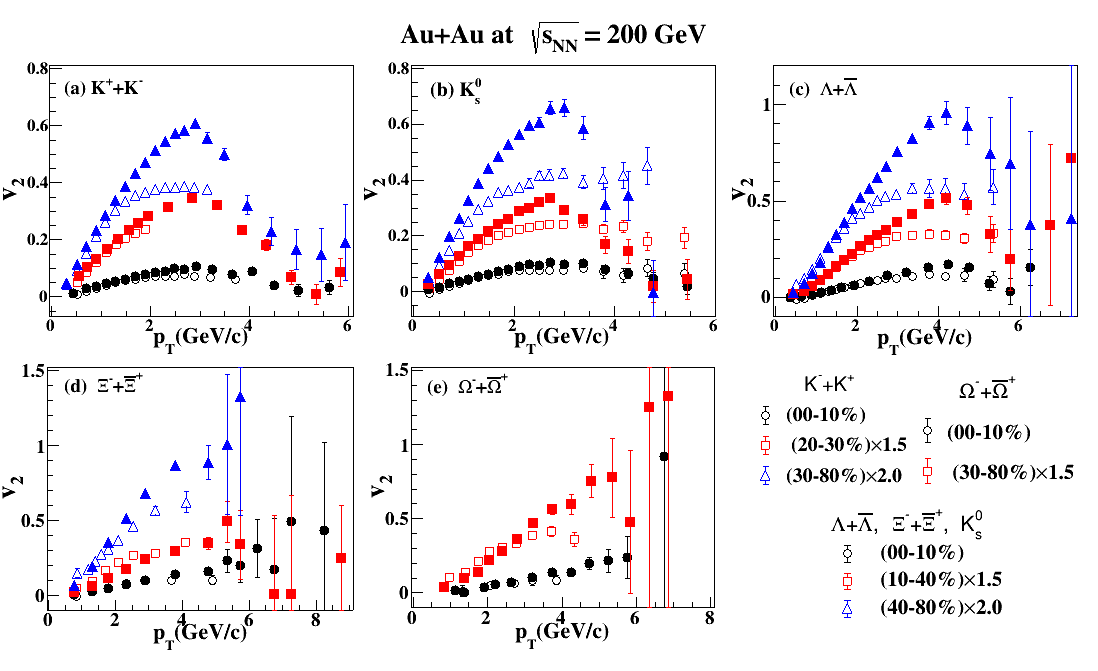}
\caption{The second harmonic coefficient $v_{2}$ of (multi-) strange hadrons as a function of transverse momentum $p_{T}$ for different centrality intervals at $\sqrt{s_{NN}}$= 200 GeV in Au+Au collisions, for (a)$K^{-}+K^{+}$, (b) $K_{s}^{0}$, (c) $\Lambda+\overline{\Lambda}$, (d) $\Xi^{-}+\overline{\Xi}^{+}$, and (e) $\Omega^{-}+\overline{\Omega}^{+}$. The solid markers represent the HYDJET++ model results while the open markers represent experimental data from the PHENIX \cite{PHENIX:2014uik}, STAR \cite{STAR:2015gge},\cite{ STAR:2008ftz} collaborations.}
\label{fig3}
\end{figure*}
\begin{figure*}[ht!]
\includegraphics[width=0.805\textwidth]{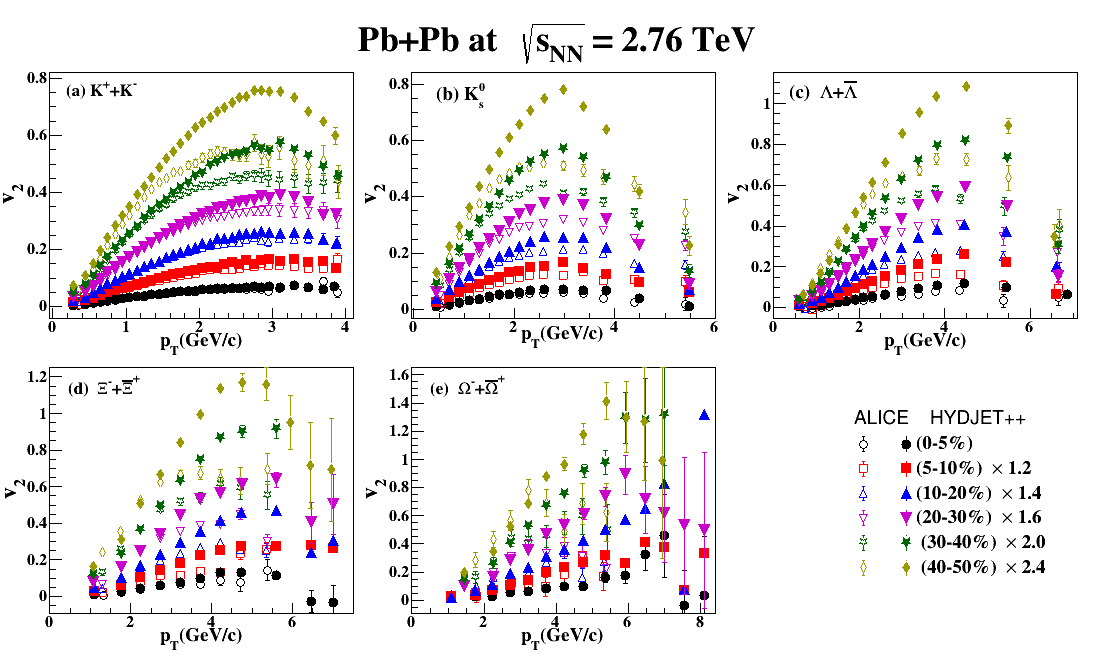}
\caption{The elliptic flow $v_{2}$ of (multi-) strange hadrons as a function of transverse momentum $p_{T}$ for different centrality intervals at $\sqrt{s_{NN}}$= 2.76 TeV in Pb+Pb collisions, for (a)$K^{-}+K^{+}$, (b) $K_{s}^{0}$, (c) $\Lambda+\overline{\Lambda}$, (d) $\Xi^{-}+\overline{\Xi}^{+}$, and (e) $\Omega^{-}+\overline{\Omega}^{+}$. The solid markers represent the HYDJET++ model results while the open markers represent experimental data from the ALICE collaboration~\cite{ALICE:2014wao}.}
\label{fig4}
\end{figure*} 

\section{Results and Discussions}\label{Results}
\subsection{Higher harmonics flow as a function $p_{T}$ at different centrality interval}
The top ((a)-(c)) row of~\Cref{fig1} shows $v_{n}$ (n=2,3,4) as a function of  $p_{T}$ for primary charged hadrons in Au+Au collisions at $\sqrt{s_{NN}}$ = 200 GeV. The results obtained by the HYDJET++ model are presented for three different centrality intervals i.e, $(0-10)\%$, $(20-30)\%$, and $(40-50)\%$. The model results are compared with the available experimental data from PHENIX collaboration~\cite{PHENIX:2011yyh}. We observe that the $v_{n}$ for charged hadrons rises linearly with $p_{T}$ initially in low $p_{T}$ region i.e., $p_{T}\leq$ 2-3 GeV/c, where a maximum is matched. With further increase in $p_{T}$, $v_{n}$ starts decreasing for all the centrality intervals in Au+Au collisions. The model calculations provide a good estimate of the experimental data~\cite{PHENIX:2011yyh} towards low $p_{T}$ region for $v_{2}$, $v_{3}$, and $v_{4}$ in all the centrality intervals. 

Similarly the bottom ((a)-(c)) row of \Cref{fig1} shows the variation of $v_{n}$ as a function of $p_{T}$ for charged hadrons in Pb+Pb collisions at $\sqrt{s_{NN}}$ = 2.76 TeV. The results are shown for three centrality intervals i.e, $(0-5)\%$, $(20-30)\%$, and $(40-50)\%$. The model results are compared with experimental data from ATLAS collaboration~\cite{ATLAS:2012at}. It is observed that the model results for $v_{2}$ and $v_{3}$ suitably match with the experimental data~\cite{ATLAS:2012at} towards low $p_{T}$ in all the centrality intervals. At high $p_{T}$, the model results underestimate the experimental data. The model results of $v_{4}$ underestimate the experimental data in central collisions because higher flow coefficients are produced by the nonlinear contributions of $v_{2}$ and $v_{3}$ thereby dominating over the intrinsic momentum anisotropy $v_{n}$. However, it provides a good estimate of the experimental data for $(20-30)\%$, and $(40-50)\%$ centrality intervals~\cite{Bravina:2016bei}. These results are well matched with the results obtained by HYDJET++ of $p_{T}$-differential $v_{3}$ and $v_{4}$ for charged hadrons in Pb+Pb collisions at $\sqrt{s_{NN}}$ = 2.76 TeV~\cite{Bravina:2013xla}. One particular observation in both Au+Au and Pb+Pb collisions is that in the intermediate $p_{T}$ region the model overestimates the experimental data of charged hadrons. This observation is discussed in detail in the article~\cite{Bravina:2013xla}, where authors suggested that it can be improved by incorporating mini-jet production or some other mechanisms. 

\begin{figure*}[hbtp]
\includegraphics[width=0.815\textwidth]{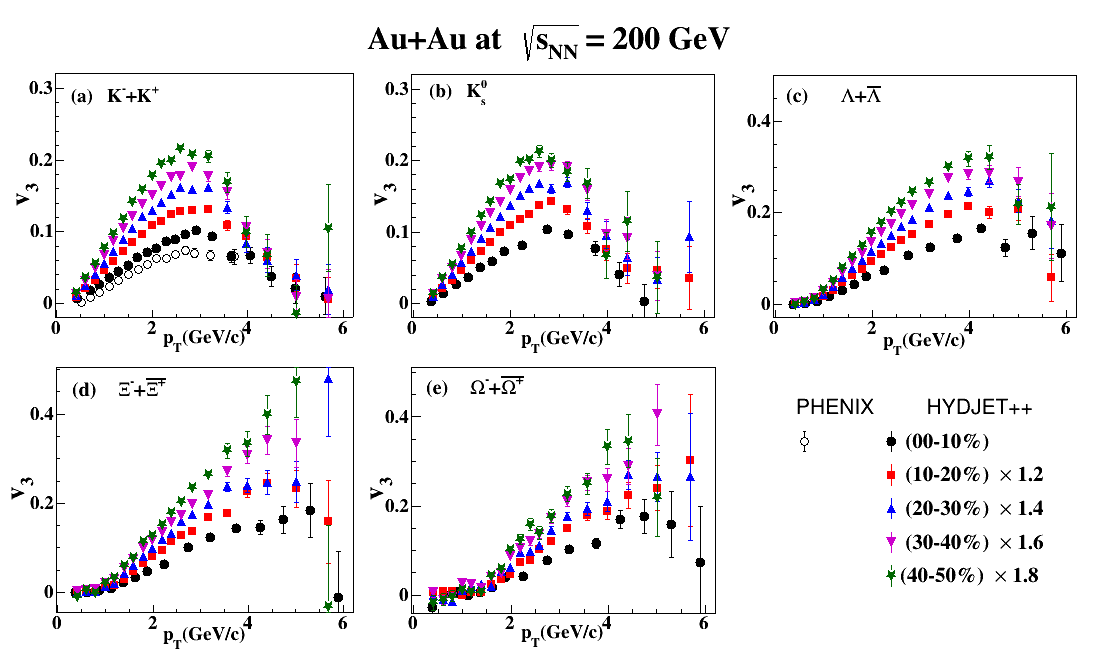}
\caption{The third harmonic coefficient $v_{3}$ of (multi-) strange hadrons as a function of $p_{T}$, for different centralities in Au+Au collisions at $\sqrt{s_{NN}}$= 200 GeV. The solid markers represent the HYDJET++ model results while the open markers represent experimental data from the PHENIX collaboration~\cite{PHENIX:2014uik}.}
\label{fig5}
\includegraphics[width=0.815\textwidth]{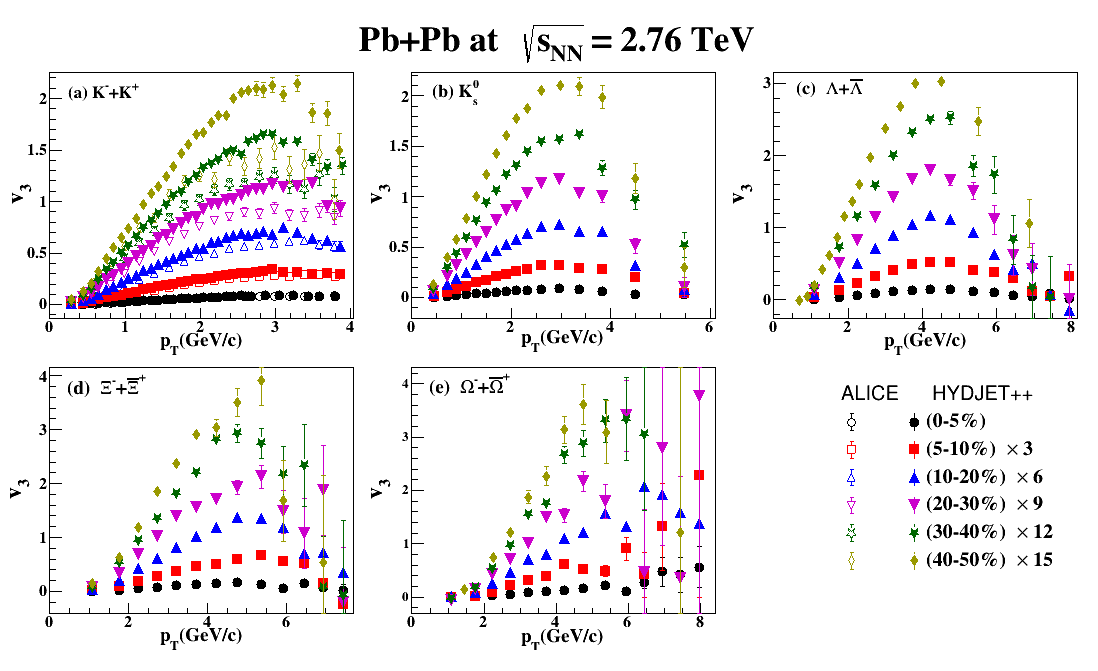}
\caption{The third harmonic coefficient $v_{3}$ of (multi-) strange hadrons as a function of $p_{T}$, for different centralities in Pb+Pb collisions at $\sqrt{s_{NN}}$= 2.76 TeV. The solid markers represent the HYDJET++ model results while the open markers represent experimental data from ALICE collaboration~\cite{ALICE:2016cti}.}
\label{fig6}
\end{figure*}

The characteristic behaviour  of $v_{3}$ and $v_{4}$ is quite similar to that of $v_{2}$ as a function of $p_{T}$ with lower magnitude. It shows that elliptic flow is a dominant source of anisotropic flow. Using the same model and with the same tuned parameter values, we study the anisotropic flow of (multi-) strange hadrons in Au+Au and Pb+Pb collision at RHIC and LHC energies, respectively.

\begin{figure*}[hbtp]
\includegraphics[width=0.815\textwidth]{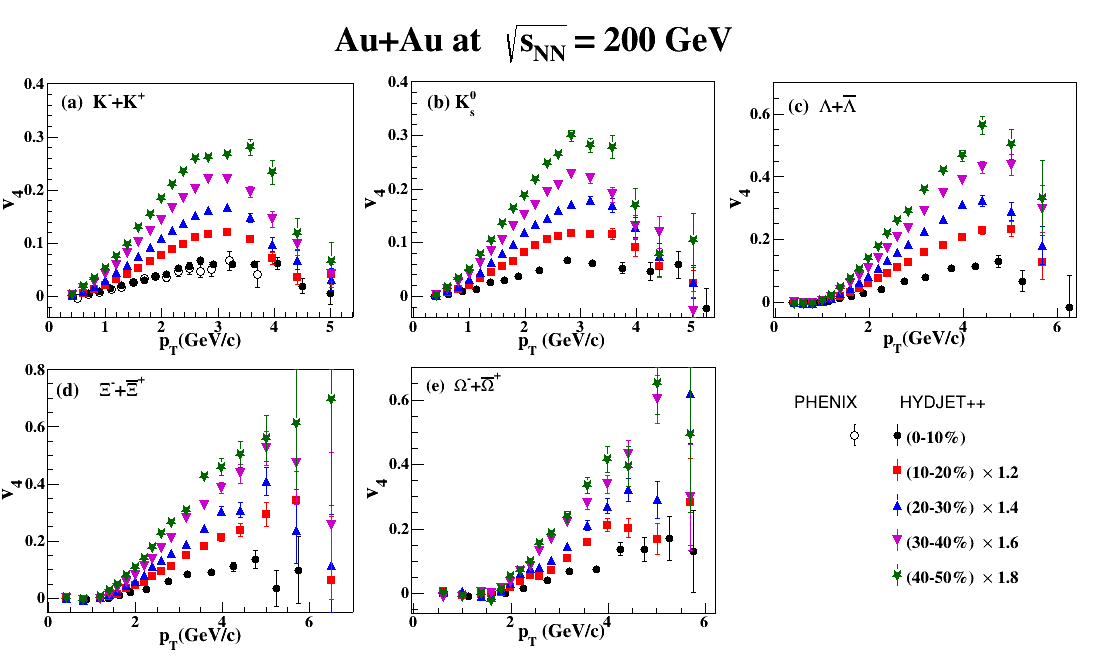}
\caption{The fourth harmonic coefficient $v_{4}$ of (multi-) strange hadrons as a function of $p_{T}$, for different centralities in Au+Au collisions at $\sqrt{s_{NN}}$= 200 GeV. The solid markers represent the HYDJET++ model results while the open markers represent experimental data from PHENIX collaboration~\cite{PHENIX:2014uik}.}
\label{fig7}
\begin{center}
\includegraphics[width=0.805\textwidth]{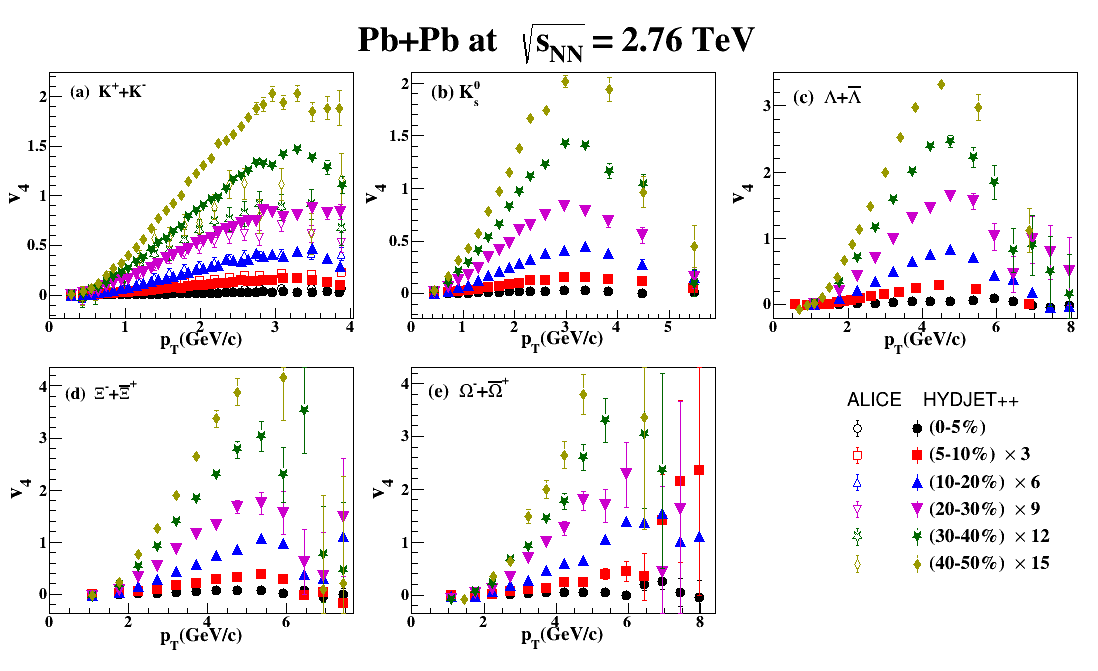}
\caption{The fourth harmonic coefficient $v_{4}$ of (multi-) strange hadrons as a function of $p_{T}$ for different centralities for Pb+Pb collisions at $\sqrt{s_{NN}}$= 2.76 TeV. The solid markers represent the HYDJET++ model results while the open markers represent experimental data from  ALICE collaboration~\cite{ALICE:2016cti}.}
\label{fig8}
\end{center}
\end{figure*}

\begin{figure*}[hbt!]
\includegraphics[width=0.805\textwidth]{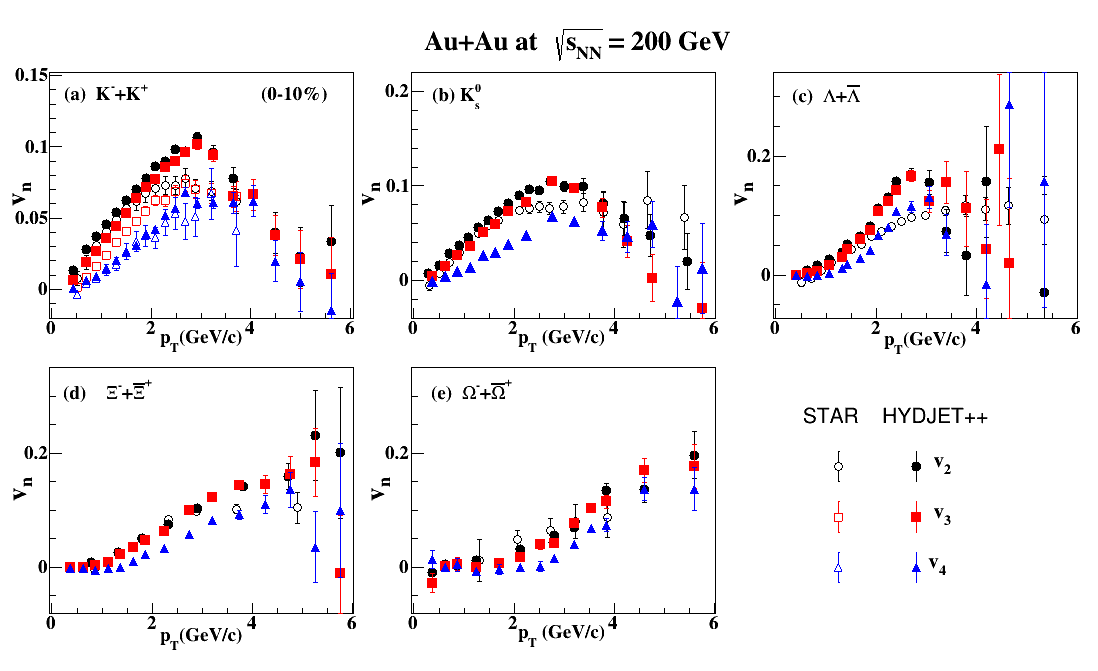}
\caption{The higher harmonic coefficients $v_{n}$ of (multi-) strange hadrons as a function of transverse momentum $p_{T}$ for $(0-10)\%$ centrality interval at $\sqrt{s_{NN}}$= 200 GeV in  Au+Au collisions.The solid markers represent the HYDJET++ model results while the open markers represent experimental data from the STAR, PHENIX collaborations\cite{PHENIX:2014uik}, \cite{STAR:2008ftz}.}
\label{fig9}
\begin{center}
\includegraphics[width=0.805\textwidth]{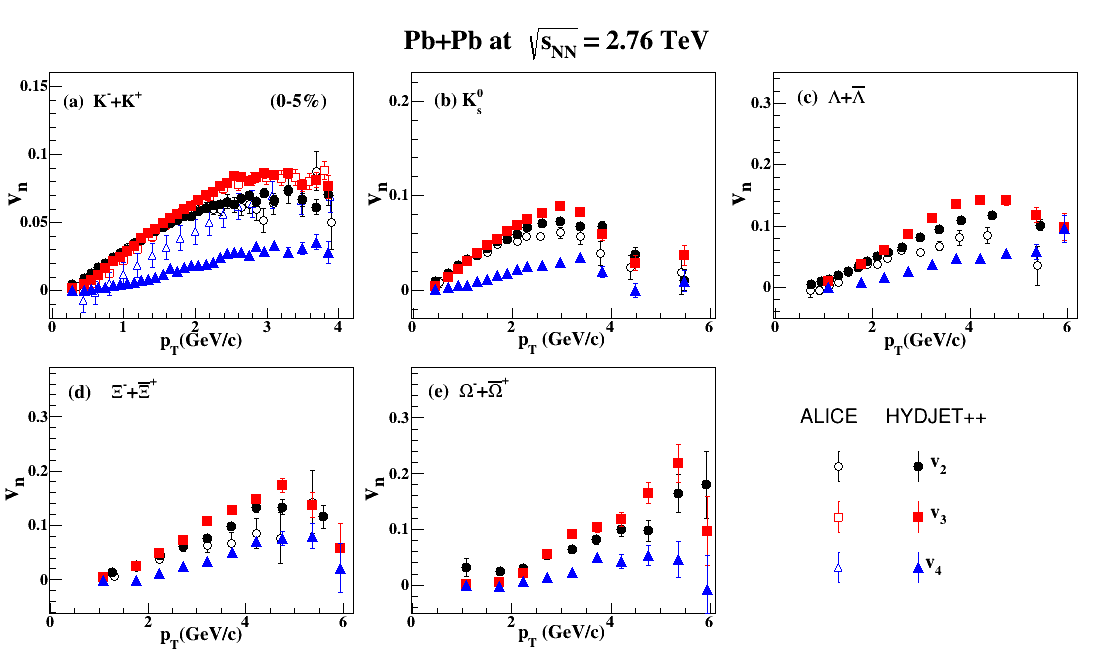}
\caption{The higher harmonic coefficients $v_{n}$ of (multi-) strange hadrons as a function of transverse momentum $p_{T}$ for central $(0-5)\%$ at $\sqrt{s_{NN}}$= 2.76 TeV in Pb + Pb collisions. The solid markers represent the HYDJET++ model results while the open markers represent experimental data from ALICE \cite{ALICE:2014wao}.}
\label{fig10}
\end{center}
\end{figure*}

\Cref{fig2} shows the variation of $v_{n}$ as a function of $p_{T}$ for (multi-) strange hadrons in Au+Au collisions at $\sqrt{s_{NN}}$ = 200 GeV obtained by HYDJET++ model. The model results are compared with the recently published experimental data~\cite{STAR:2022ncy} from STAR collaboration. The results are shown for $(0-80)\%$ centrality interval. \Cref{fig2}(a) shows that $v_{2}$ results are well matched with the experimental data~\cite{STAR:2008ftz} upto $p_{T}\leq$ 2 GeV/c and then overpredicts the experimental data with the increasing $p_{T}$. \Cref{fig2}(b) presents $v_{3}$ results for different strange hadron species. It is observed that the model provides a good description of experimental data~\cite{STAR:2022ncy} of $K^{-}+K^{+}$ and $\Lambda+\overline{\Lambda}$ upto $p_{T}\leq$ 2.5 GeV/c thereafter it overpredicts the experimental data at higher $p_{T}$. Due to non availability of experimental data for $\Xi^{-}+\overline{\Xi}^{+}$ and $\Omega^{-}+\overline{\Omega}^{+}$ in $(0-80)\%$ centrality interval, the $\Xi^{-}+\overline{\Xi}^{+}$ and $\Omega^{-}+\overline{\Omega}^{+}$ results are the model predictions. We expect that it should follow the trends of $K^{-}+K^{+}$ and $\Lambda+\overline{\Lambda}$. \Cref{fig2}(c) shows the results of $v_{4}$ for strange hadrons. It is observed that the model overpredicts the experimental data of $K^{-}+K^{+}$. The model prediction for $v_{4}$ of  $K_{s}^{0}$, $\Lambda+\overline{\Lambda}$, $\Xi^{-}+\overline{\Xi}^{+}$ and $\Omega^{-}+\overline{\Omega}^{+}$ are also presented. It is observed that the model describes the experimental result of a rapid increase in $v_{n}$ of mesons ($K^{-}+K^{+}$ and $K_{s}^{0}$) in comparison to baryons at low $p_{T}$ i.e, $p_{T}\leq$ 1 GeV/c. It is because of the fact that the probability of formation of $K^{-}$ mesons by a combination of a strange quark and an up or a down quark is much higher than the baryon formation at low $p_{T}$.

\Cref{fig3,fig4} show elliptic flow $v_{2}(p_{T})$ results obtained from HYDJET++ model for $K^{-}+K^{+}$, $K_{s}^{0}$, $\Lambda+\overline{\Lambda}$, $\Xi^{-}+\overline{\Xi}^{+}$ and $\Omega^{-}+\overline{\Omega}^{+}$ in various centrality intervals for Au+Au collisions at $\sqrt{s_{NN}}$ = 200 GeV and Pb+Pb collisions at $\sqrt{s_{NN}}$ = 2.76 TeV, respectively. The model results are compared with the corresponding experimental data from STAR, PHENIX, and ALICE collaborations~\cite{STAR:2015gge, PHENIX:2014uik, STAR:2008ftz, ALICE:2014wao}. It is observed that the model reproduces the experimental data of (multi-) strange hadrons. However, towards peripheral collisions the match between the model results and experimental data is limited to much lower $p_{T}$ regions. This may be due to the fact that the model parameters in the present study responsible for the evaluation of anisotropic flow coefficient is same for all the centralities. Fine-tuning of these parameters for different centrality intervals may lead to better agreement between the model results and experimental data. It is also observed that moving from central to peripheral collisions, maxima of $v_{2}$ increases and shifts towards lower $p_{T}$. It is because of the smaller radial flow generated during fireball expansion in peripheral collisions compared to that produced in central collisions~\cite{ALICE:2014wao}.

In the same way, \Cref{fig5,fig6} show variation of $v_{3}$ as a function of $p_{T}$ for (multi-) strange hadrons obtained by HYDJET++ model for different centrality intervals in Au+Au and Pb+Pb collisions, respectively. The model results are compared  with the available experimental data from PHENIX and ALICE experiments~\cite{PHENIX:2014uik, ALICE:2016cti}. The Au+Au results  for $v_{3}$ are similar to those described in \Cref{fig2} and Pb+Pb results show same behaviour as $v_{2}$ in \Cref{fig4} but with different magnitude.

 Likewise, \Cref{fig7,fig8} show the variation of $v_{4}$ as a function of $p_{T}$ in Au+Au and Pb+Pb collisions, respectively, for different centrality intervals. We observe a much better agreement between $v_{4}$ results obtained from the model and the experimental data towards central collisions in comparison to $v_{3}$ results.
\begin{figure*}[hbtp]
\begin{center}
\includegraphics[width=0.901\textwidth]{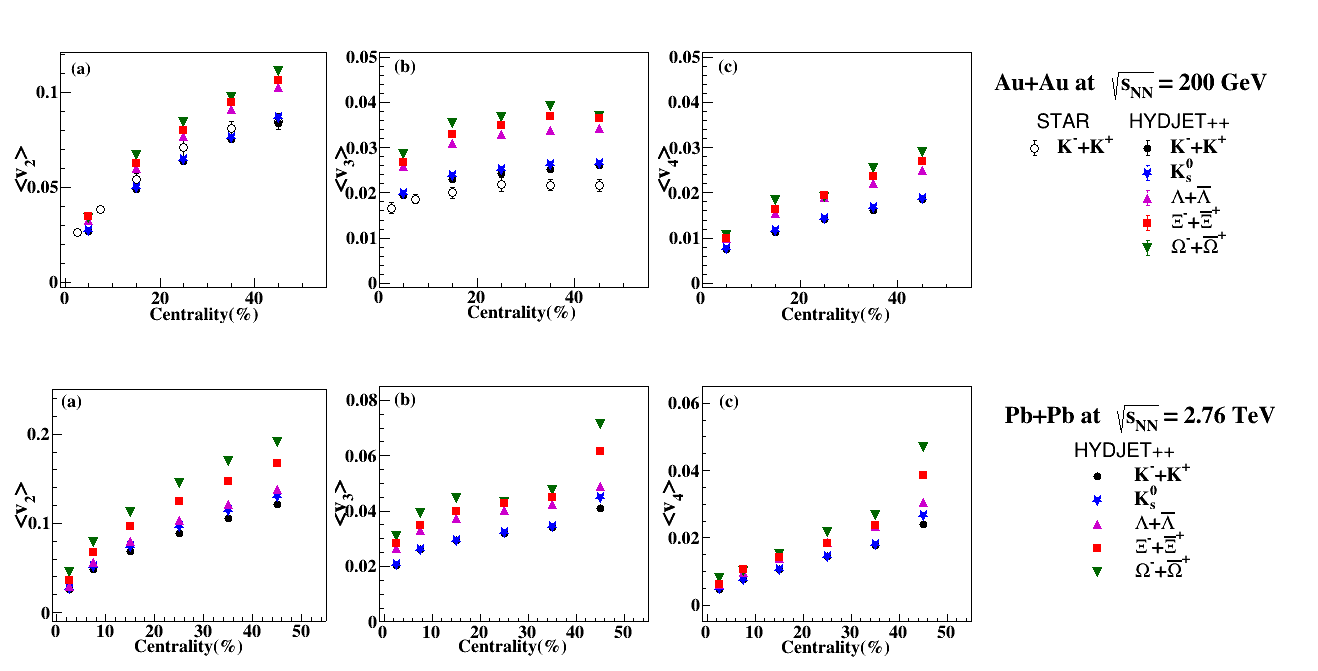}
\caption{$p_{T}$ integrated flow  $v_{2}$, $v_{3}$, and $v_{4}$ of (multi-) strange hadrons as a function of centrality in Au+Au collisions at $\sqrt{s_{NN}}$= 200 GeV and $\sqrt{s_{NN}}$= 2.76 TeV in Pb+Pb collisions. The solid markers represent the HYDJET++ model results while the open markers represent experimental data from STAR~\cite{STAR:2022ncy} .}
\label{fig11}
\end{center}
\end{figure*}

\Cref{fig9,fig10} show a comparison among $v_{2}$, $v_{3}$, and $v_{4}$ of (multi-) strange hadrons in most central collision i.e, $(0-10)\%$ centrality interval for Au+Au and $(0-5)\%$ centrality interval for Pb+Pb collisions, respectively. It is observed that HYDJET++ elliptic flow $v_{2}$ has slightly higher magnitude than $v_{3}$ at low $p_{T}$ i.e, $p_{T}\leq$ 2 GeV/c in Au+Au collisions for all the strange hadron species. At higher $p_{T}$, both $v_{2}$ and $v_{3}$ have same magnitude. However, in Pb+Pb collisions, $v_{2}$ is slightly higher in magnitude than $v_{3}$ up to $p_{T}\leq$ 1.5 GeV/c, beyond which $v_{3}$ is larger than $v_{2}$. It indicates that event-by-event fluctuations contribute more to the development of flow in most central collisions~\cite{PHENIX:2014uik, STAR:2008ftz, ALICE:2014wao, ALICE:2016cti, ATLAS:2018ezv} which gets reflected in the higher value of $v_{3}$ than $v_{2}$ at intermediate $p_{T}$. It can be seen that $v_{4}$ for $K^{-}+K^{+}$ matches well with the experimental data in Au+Au collisions. For Pb+Pb collisions, the model underpredicts the experimental data of $v_{4}$ for $K^{-}+K^{+}$. This observation is in accordance with the results obtained in ref.~\cite{Bravina:2013xla} for Pb+Pb collisions at $\sqrt{s_{NN}}$ = 2.76 TeV.  

\subsection{ Higher harmonics flow as a function of centrality}
\Cref{fig11} shows $p_{T}$ integrated $v_{n}$ as a function of centrality for (multi-) strange hadrons in Au+Au collisions at $\sqrt{s}_{NN}$ =200 GeV and Pb+Pb collisions at  $\sqrt{s}_{NN}$ =2.76 TeV, respectively. For comparision, the available experimental data of $K^{+}+K^{-}$ from STAR collaboration~\cite{STAR:2022ncy} is also shown in  top row of \Cref{fig11}. The model approximately reproduces the experimental $\left<v_{2}\right>$ data for $K^{+}+K^{-}$ but it underestimates the $\left<v_{3}\right>$ data for all the centrality intervals. It is observed that $\left<v_{2}\right>$, $\left<v_{3}\right>$ and $\left<v_{4}\right>$ increase with increasing centrality. $\left<v_{3}\right>$ initially increases with increasing centrality but towards peripheral collision, a saturation is observed in $\left<v_{3}\right>$ value. It may happen because event-by-event fluctuations dominate in $v_{3}$ in central collisions. Towards peripheral collisions, the contribution of fluctuations is small and shows weak dependence on centrality since odd-order flow harmonics are produced only by fluctuations~\cite{Bravina:2013xla, Solanki:2012ne}. 

\begin{figure*}[hbtp]
\begin{center}
\includegraphics[width=0.81\textwidth]{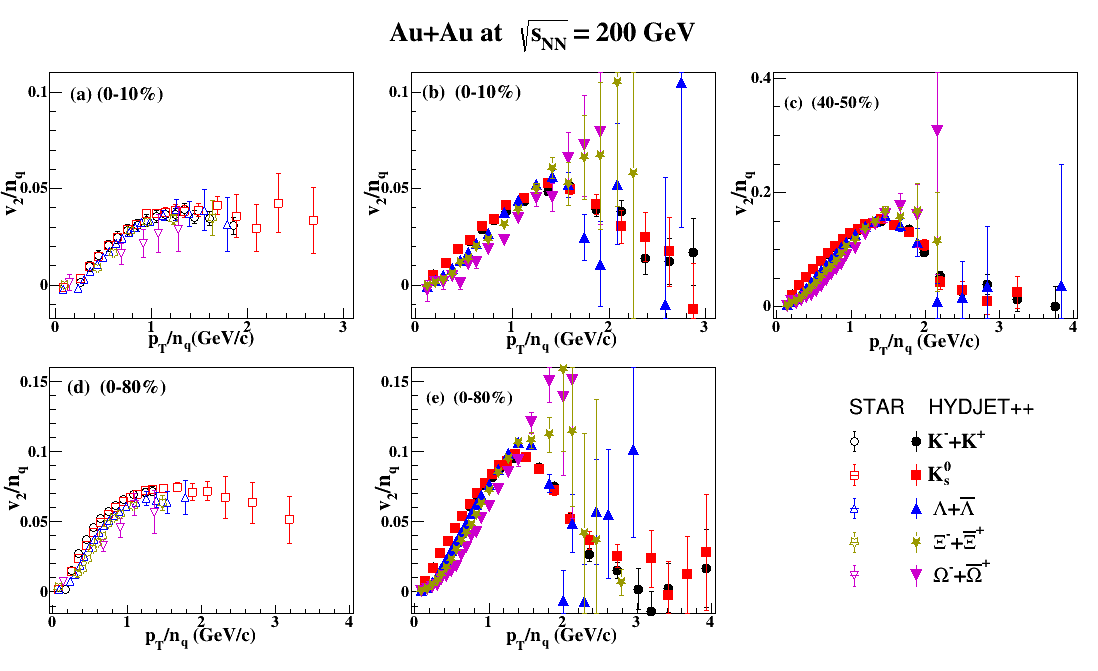}
\caption{The scaled elliptic flow $v_{2}/n_{q}$ versus the scaled transverse momentum $p_{T}/n_{q}$ of (multi-) strange hadrons  for $(0-10)\%$, and $(0-80)\%$ centrality intervals in Au+Au collisions at $\sqrt{s_{NN}}$= 200 GeV. The solid markers represent the HYDJET++ model results while the open markers represent experimental data from STAR and PHENIX collaborations~\cite{STAR:2022ncy, STAR:2008ftz, PHENIX:2014uik}.}
\label{fig12}
\end{center}
\end{figure*}

From top row of \Cref{fig11}, baryon-meson grouping is clearly observed in $\left<v_{n}\right>$ for Au+Au collisions at $\sqrt{s_{NN}}$ =200 GeV, but it is absent in Pb+Pb collisions at $\sqrt{s_{NN}}$ =2.76 TeV in bottom row of \Cref{fig11}. This is because of the large contribution of jets at LHC energy regime. It is the main motivation to test NCQ scaling at both RHIC and LHC energies.

\begin{figure*}[hbt!]
\begin{center}
\includegraphics[width=0.82\textwidth]{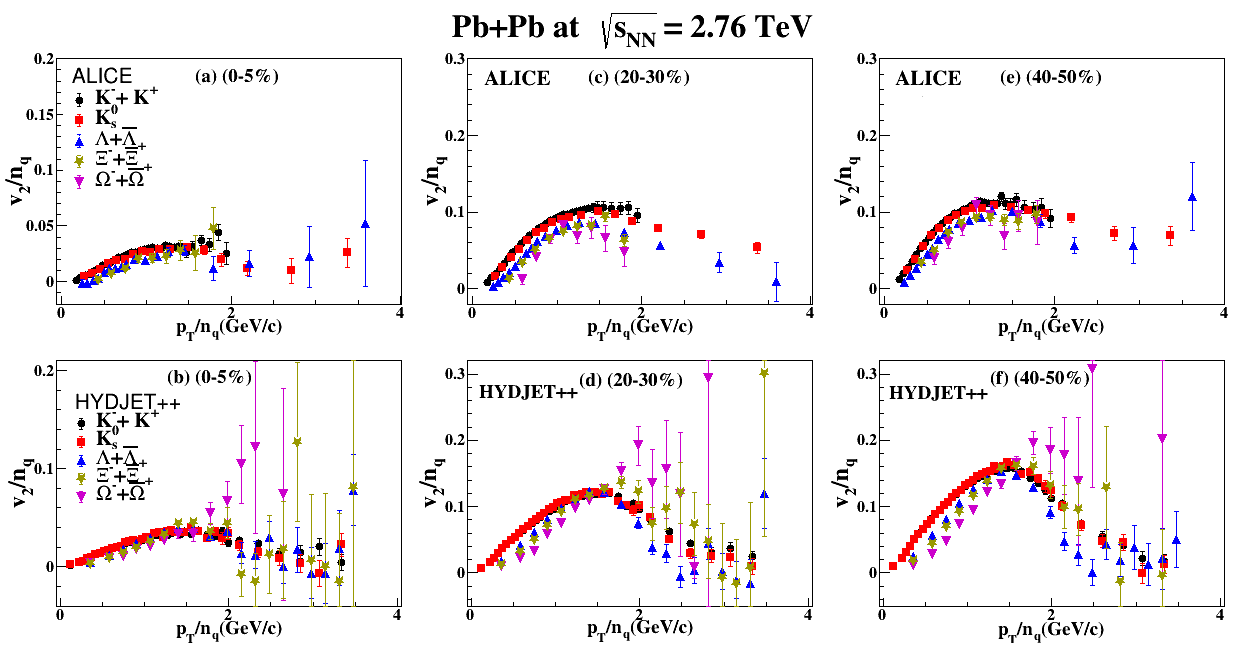}
\caption{The scaled  elliptic flow $v_{2}/n_{q}$ versus the scaled transverse momentum $p_{T}/n_{q}$  of (multi-) strange hadrons for $(0-5)\%$, $(20-30)\%$ and $(40-50)\%$ centrality intervals in Pb+Pb collisions at $\sqrt{s_{NN}}$= 2.76 TeV \cite{ALICE:2014wao}.}
\label{fig13}
\end{center}
\end{figure*}
\begin{figure*}[hbt!]
\begin{center}
\includegraphics[width=0.829\textwidth]{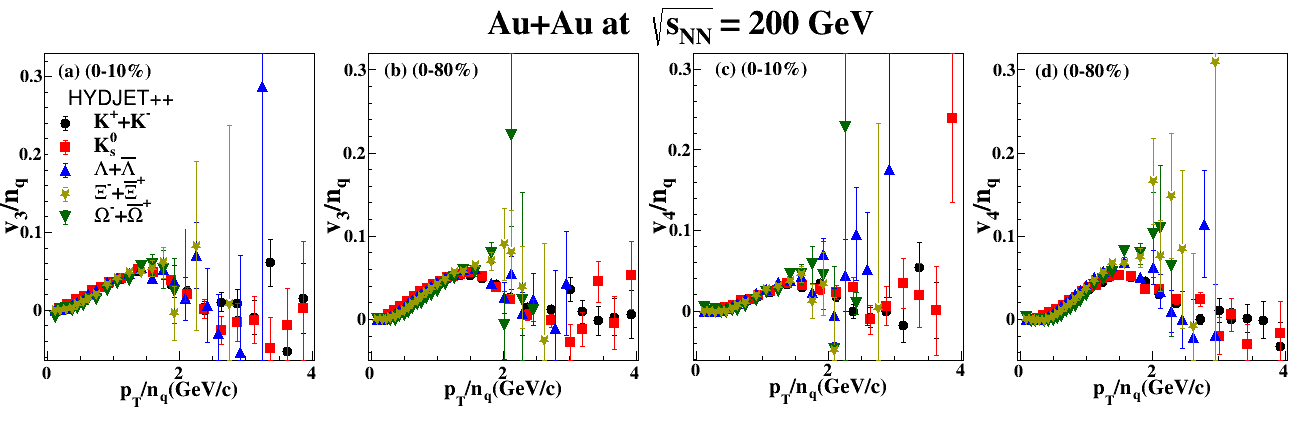}
\caption{The scaled  triangular $v_{3}/n_{q}$ and quadrangular $v_{4}/n_{q}$ flows versus the scaled transverse momentum $p_{T}/n_{q}$ of (multi-) strange hadrons for $(0-10)\%$ and $(0-80)\%$ centrality intervals in Au+Au collisions at $\sqrt{s_{NN}}$= 200 GeV.}
\label{fig14}
\end{center}
\end{figure*}

\begin{figure*}[hbt!]
\begin{center}
\includegraphics[width=0.81\textwidth]{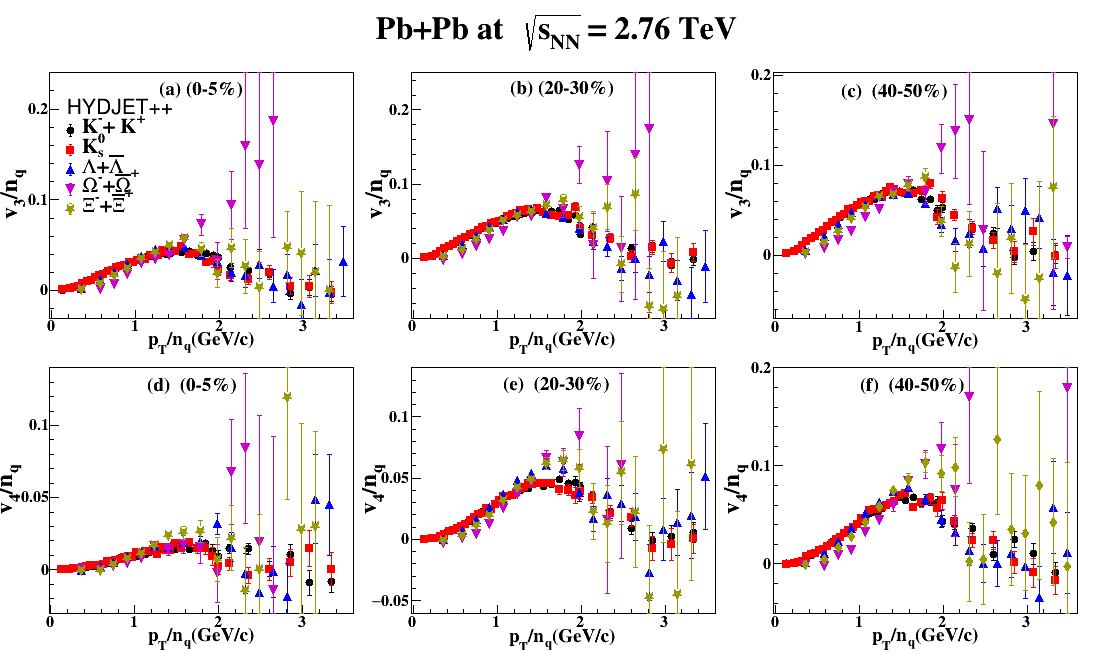}
\caption{The scaled  triangular $v_{3}/n_{q}$ and quadrangular $v_{4}/n_{q}$ flows versus the scaled transverse momentum $p_{T}/n_{q}$ of (multi-) strange hadrons for $(0-5)\%$, $(20-30)\%$, and $(40-50)\%$ centrality intervals in Pb+Pb collisions at $\sqrt{s_{NN}}$= 2.76 TeV.}
\label{fig15}
\end{center}
\end{figure*}
\subsection{Test of NCQ scaling}
\Cref{fig12} shows NCQ scaling results in Au+Au collisions at $\sqrt{s_{NN}}$ =200 GeV obtained from HYDJET++ model and compared with corresponding results from STAR collaboration~\cite{STAR:2022ncy, STAR:2008ftz, PHENIX:2014uik}. STAR data is shown in \Cref{fig12}(a) for $(0-10)\%$ centrality interval and in \Cref{fig12}(d) for $(0-80)\%$ interval. HYDJET++ results on NCQ scaling are presented in \Cref{fig12}(b) and \Cref{fig12}(e) for $(0-10)\%$ and $(0-80)\%$ centrality intervals, respectively. 
\begin{figure*}[hbtp]
\begin{center}
\includegraphics[width=0.947\textwidth]{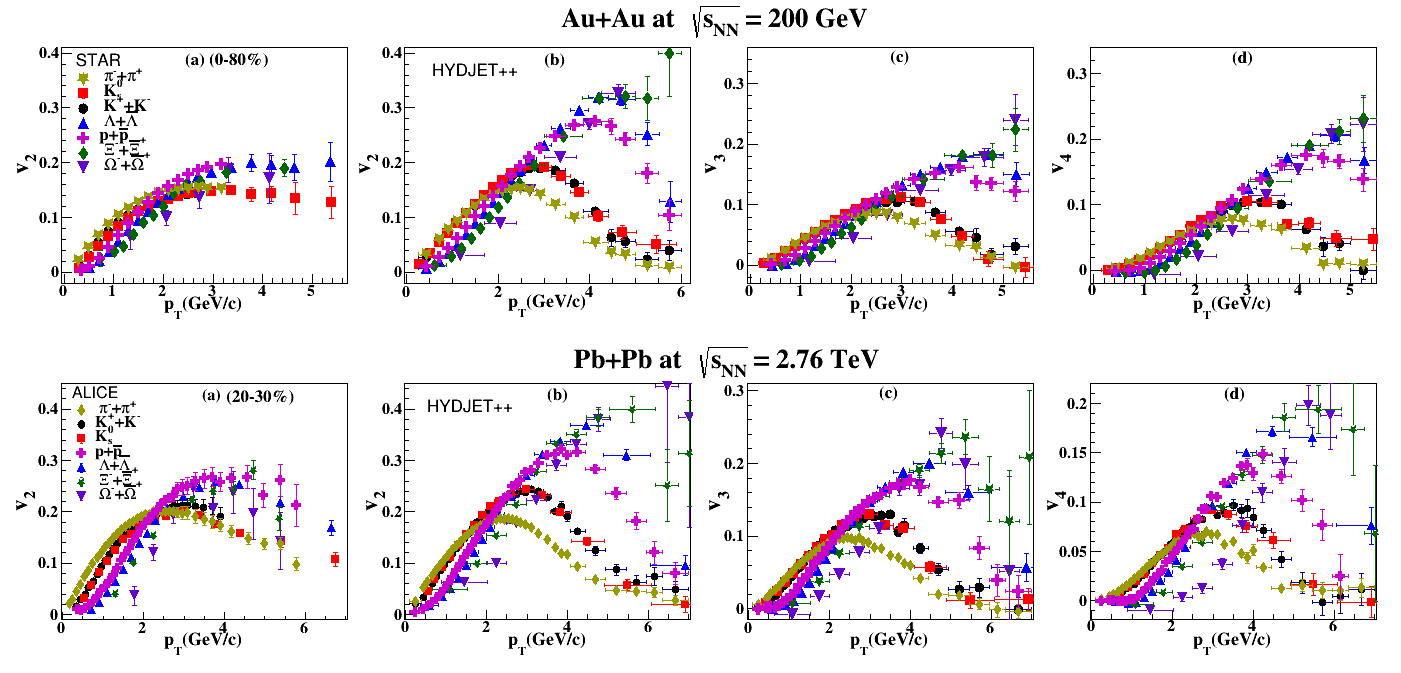}
\caption{$v_{n}$ of (non-) strange hadrons as a function $p_{T}$ for $(0-80)\%$ and $(20-30)\%$ centrality intervals in Au+Au collisions at $\sqrt{s_{NN}}$= 200 GeV and in Pb+Pb collisions at $\sqrt{s_{NN}}$= 2.76 TeV, respectively. Experimental results from STAR~\cite{STAR:2008ftz, STAR:2015gge, STAR:2022ncy} and ALICE~\cite{Zhu:2016qiv, ALICE:2014wao} collaborations for elliptic flow $v_{2}$ are shown in panel (a), while HYDJET++ results for all $v_{n}$ are shown in (b), (c) and (d).}
\label{fig16}
\end{center}
\end{figure*}

The NCQ scaling is studied by the variation of $v_{2}/n_{q}$ as a function of $p_{T}/n_{q}$, where $n_{q}$ is the number of constituent quark. It is observed that the NCQ scaling remains valid for strange mesons throughout the $p_{T}$ range in $(0-10)\%$ centrality. For strange baryons, the NCQ scaling is valid up to $p_{T}\approx$ 1.5 GeV. Thereafter, because of large fluctuation in model calculation for strange baryons, the NCQ scaling can not be described properly. However, increasing the event statistics may reduce the large fluctuation in strange baryon results. Comparing the HYDJET++ results with the experimental data in \Cref{fig12}(a) and (b), it is observed that the model calculations reproduce the shape of the experimental data for all the particle species but with higher magnitude. Similar observations are made in \Cref{fig12}(d) and (e). In \Cref{fig12}(c), for non-central collisions, it is observed that the strange baryon results are consistently below the strange meson results. It is because of the transition between thermal and thermal+shower recombination. For peripheral collisions, the transition can happen at lower $p_{T}/n_{q}$ while in central collisions, this transition occurs at higher $p_{T}/n_{q}$ value~\cite{PHENIX:2012swz}. 

Likewise, \Cref{fig13} shows the results of NCQ scaling in Pb+Pb collisions at $\sqrt{s_{NN}}=$ 2.76 TeV. The experimental data and the model results are shown separately for $(0-5)\%$, $(20-30)\%$, and $(40-50)\%$ centrality intervals, respectively. The observations are similar to that of NCQ scaling in Au+Au collisions. However, in non-central collisions a clear separation between strange meson and baryon spectra is observed  in both the model calculations and experimental data. It indicates the violation of NCQ scaling in non-central collisions for Pb+Pb collisions at LHC~\cite{ALICE:2018yph}.

\Cref{fig14,fig15} show NCQ scaling of $v_{3}$ and $v_{4}$ in HYDJET++ for Au+Au and Pb+Pb collisions, respectively over different centrality intervals. Here also, it can be seen that the model clearly shows the NCQ scaling for all the hadrons in Au+Au collisions up to $p_{T}\approx$ 1.5 GeV, while for Pb+Pb collisions a violation in the NCQ scaling is observed towards peripheral collisions due to stronger influence of jet hadrons at LHC~\cite{Crkovska:2016flo, ALICE:2018yph}.

\subsection{ Mass Ordering effect in flow harmonics of (non-)strange hadrons} 

 \Cref{fig16} presents the mass ordering of anisotropic flow among $\pi^{+}+\pi^{-}$, $K^{+}+K^{-}$, $K_{s}^{0}$, $p+ \overline{p}$, $\Lambda+\overline{\Lambda}$, $\Xi^{-}+\overline{\Xi}^{+}$, and $\Omega^{-}+\overline{\Omega}^{+}$) in Au+Au collisions at $\sqrt{s}_{NN}$ =200 GeV and Pb+Pb collisions at $\sqrt{s}_{NN}$ =2.76 TeV, respectively. The available experimental data and the model results for $v_{2}$ are shown separately. The results are shown for $(0-80)\%$ centrality interval in Au+Au collisions and for $(20-30)\%$ in Pb+Pb collisions. It can be seen that the model reproduces the experimental data of $v_{2}$ up to $p_{T}\approx$ 2.0 GeV/c implying that the model provides a good description of mass ordering among all the hadron species at low $p_{T}$. This is because of the interplay between radial and elliptic flows generated during hadronic evolution. The radial flow pushes the heavier particles towards higher $p_{T}$ which gets manifested to $p_{T}$-dependent mass ordering of elliptic flow at low $p_{T}$. This tends to decrease the elliptic flow with increase of hadron mass. Towards higher $p_{T}$, baryon-meson grouping (particle-type dependence) is observed in the model calculation and in the experimental data. From the figures, we also observe that the separation between mass ordering of $\pi^{+}+\pi^{-}$ and $K^{+}+K^{-}$ is less visible in HYDJET++ model in comparison to ALICE experimental data. However, experimental results show a clear separation between the mass ordering for all the hadrons species. This may be due to the strangeness enhancement factor $\gamma_{s}$ used in soft particle production component of the model. Similar observations are made for $v_{3}$ and $v_{4}$ in Au+Au and Pb+Pb collisions. 

\section{Conclusion}
\label{conclusion}
We have presented $p_{T}$-differential flow harmonics $v_{n}(n=2,3,4)$ of (multi-) strange hadrons at RHIC and LHC energies using HYDJET++ model. We have compared our findings with the available experimental data at both energies. HYDJET++ model suitably describes the experimental data at low $p_{T}$. At intermediate $p_{T}$, the model overpredicts the experimental data for $v_{n}$. The difference between the model calculations and the experimental data increases as we move toward peripheral collisions. This may be because the mechanism involved to address the interaction between the medium and jet partons is absent in the HYDJET++ model framework at the present scenario. We observe a clear baryon-meson grouping in $p_{T}$-integrated flow coefficients $v_{n}$ in Au+Au collisions at $\sqrt{s}_{NN}$= 200 GeV. The meson-baryon grouping is absent in Pb+Pb collisions at $\sqrt{s}_{NN}$= 2.76 TeV. 

We observed that the model reproduces the experimental results on constituent quark scaling at both RHIC and LHC energies. Moreover, similar to the experimental results, a violation of NCQ scaling is also observed in the model calculations at LHC energy. Further, we also observed mass ordering in the anisotropic flow of strange and non-strange hadrons at low $p_{T}$ in both Au+Au and Pb+Pb collisions, where mesons have higher magnitude of flow than the baryons. At intermediate $p_{T}$, the mass order reverses, and heavier baryons acquire higher flow in comparison to mesons.

Thus, our study  highlighted the contribution of flow of (multi-) strange hadrons to the collective properties of QGP medium occurring in the partonic phase of Au+Au and Pb+Pb collision systems at RHIC and LHC energy regimes, respectively. Comparison of the $v_{2}$($p_{T}$), $v_{3}$($p_{T}$), and $v_{4}$($p_{T}$) values for different particle species in these two different energy regimes might be helpful in providing additional insight into the dynamics of anisotropic flow and the effect of radial expansion of the system.
 
\section{Acknowledgements}
BKS gratefully acknowledges the financial support provided by the BHU Institutions of Eminence (IoE) Grant No. 6031, Govt. of India. AS would like to thank CSIR, India for providing Senior Research Fellowship. SP and GD acknowledge the financial support obtained from UGC under the research fellowship scheme in central universities.


\begin{thebibliography}{100}
\bibitem{ALICE:2021ibz}S.~Acharya \textit{et al.} [ALICE], JHEP \textbf{10} (2021), 152 [arXiv:2107.10592 [nucl-ex]].
\bibitem{Snellings:2011sz} R.~Snellings, New J. Phys. \textbf{13} (2011), 055008 [arXiv:1102.3010 [nucl-ex]].
\bibitem{Snellings:2014vqa}R.~Snellings, EPJ Web Conf. \textbf{97} (2015), 00025 [arXiv:1411.7690 [nucl-ex]]. 
\bibitem{STAR:2021twy} M.~Abdallah \textit{et al.} [STAR], Phys. Rev. C \textbf{103} (2021) no.6, 064907 [arXiv:2103.09451 [nucl-ex]].
\bibitem{STAR:2005gfr}J.~Adams \textit{et al.} [STAR], Nucl. Phys. A \textbf{757} (2005), 102-183 [arXiv:nucl-ex/0501009 [nucl-ex]].
\bibitem{Voloshin:2008dg} S.~A.~Voloshin, A.~M.~Poskanzer and R.~Snellings, Landolt-Bornstein \textbf{23} (2010), 293-333 [arXiv:0809.2949 [nucl-ex]].
\bibitem{Voloshin:1994mz} S.~Voloshin and Y.~Zhang, Z. Phys. C \textbf{70} (1996), 665-672 [arXiv:hep-ph/9407282 [hep-ph]].
\bibitem{Singh:1992sp} C.~P.~Singh,
Phys. Rept. \textbf{236} (1993), 147-224
\bibitem{Stoecker:2004qu} H.~Stoecker,
Nucl. Phys. A \textbf{750} (2005), 121-147 [arXiv:nucl-th/0406018 [nucl-th]].
\bibitem{STAR:2013qio} L.~Adamczyk \textit{et al.} [STAR], Phys. Rev. C \textbf{88} (2013) no.1, 014904 [arXiv:1301.2187 [nucl-ex]].
\bibitem{Solanki:2012ne}  D.~Solanki, P.~Sorensen, S.~Basu, R.~Raniwala and T.~K.~Nayak, Phys. Lett. B \textbf{720} (2013), 352-357 [arXiv:1210.0512 [nucl-ex]].
\bibitem{ALICE:2016cti} J.~Adam \textit{et al.} [ALICE], JHEP \textbf{09} (2016), 164 [arXiv:1606.06057 [nucl-ex]].
\bibitem{STAR:2010ico} B.~I.~Abelev \textit{et al.} [STAR], Phys. Rev. C \textbf{81} (2010), 044902 [arXiv:1001.5052 [nucl-ex]].
\bibitem{Heinz:2013bua} U.~Heinz, Z.~Qiu and C.~Shen, Phys. Rev. C \textbf{87} (2013) no.3, 034913 [arXiv:1302.3535 [nucl-th]].
\bibitem{Bravina:2015sda} L.~V.~Bravina, E.~S.~Fotina, V.~L.~Korotkikh, I.~P.~Lokhtin, L.~V.~Malinina, E.~N.~Nazarova, S.~V.~Petrushanko, A.~M.~Snigirev and E.~E.~Zabrodin, Eur. Phys. J. C \textbf{75} (2015) no.12, 588 [arXiv:1509.02692 [hep-ph]].
\bibitem{Tiwari:1997zu}  V.~K.~Tiwari and C.~P.~Singh, Phys. Lett. B \textbf{411} (1997), 225-229.
\bibitem{Bazavov:2014xya} A.~Bazavov, H.~T.~Ding, P.~Hegde, O.~Kaczmarek, F.~Karsch, E.~Laermann, Y.~Maezawa, S.~Mukherjee, H.~Ohno and P.~Petreczky, \textit{et al.} Phys. Rev. Lett. \textbf{113} (2014) no.7, 072001 [arXiv:1404.6511 [hep-lat]].
\bibitem{PhysRevC.107.024906} Arpit~Singh, P.~K.~Srivastava, Gauri~Devi, and B. K. Singh, Phys. Rev. C \textbf{107} (2023), 024906.
\bibitem{STAR:2015gge} L.~Adamczyk \textit{et al.} [STAR], Phys. Rev. Lett. \textbf{116} (2016) no.6, 062301 [arXiv:1507.05247 [nucl-ex]].
\bibitem{STAR:2022ncy} M.~Abdallah \textit{et al.} [STAR], Phys. Rev. C \textbf{105} (2022) no.6, 064911 [arXiv:2203.07204 [nucl-ex]].
\bibitem{ALICE:2014wao}  B.~B.~Abelev \textit{et al.} [ALICE], JHEP \textbf{06} (2015), 190 [arXiv:1405.4632 [nucl-ex]].
\bibitem{STAR:2005npq} J.~Adams \textit{et al.} [STAR], Phys. Rev. Lett. \textbf{95} (2005), 122301 [arXiv:nucl-ex/0504022 [nucl-ex]].
\bibitem{STAR:2008ftz} B.~I.~Abelev \textit{et al.} [STAR], Phys. Rev. C \textbf{77} (2008), 054901 [arXiv:0801.3466 [nucl-ex]].
\bibitem{STAR:2022tfp}  M.~Abdallah \textit{et al.} [STAR], Phys. Rev. C \textbf{107} (2023) no.2, 024912 [arXiv:2205.11073 [nucl-ex]].
\bibitem{ALICE:2018rtz}  S.~Acharya \textit{et al.} [ALICE], JHEP \textbf{07} (2018), 103 [arXiv:1804.02944 [nucl-ex]].
\bibitem{Bravina:2013xla} L.~V.~Bravina, B.~H.~Brusheim Johansson, G.~K.~Eyyubova, V.~L.~Korotkikh, I.~P.~Lokhtin, L.~V.~Malinina, S.~V.~Petrushanko, A.~M.~Snigirev and E.~E.~Zabrodin, Eur. Phys. J. C \textbf{74} (2014) no.3, 2807 [arXiv:1311.7054 [nucl-th]].
\bibitem{Lokhtin:2012re} I.~P.~Lokhtin, A.~V.~Belyaev, L.~V.~Malinina, S.~V.~Petrushanko, E.~P.~Rogochaya and A.~M.~Snigirev, Eur. Phys. J. C \textbf{72} (2012), 2045 [arXiv:1204.4820 [hep-ph]].
\bibitem{ALICE:2018yph} S.~Acharya \textit{et al.} [ALICE],JHEP \textbf{09} (2018), 006 [arXiv:1805.04390 [nucl-ex]].
\bibitem{ALICE:2016ccg} J.~Adam \textit{et al.} [ALICE], Phys. Rev. Lett. \textbf{116} (2016) no.13, 132302 [arXiv:1602.01119 [nucl-ex]].
\bibitem{STAR:2004jwm} J.~Adams \textit{et al.} [STAR], Phys. Rev. C \textbf{72} (2005), 014904 [arXiv:nucl-ex/0409033 [nucl-ex]].
\bibitem{PHENIX:2014uik} A.~Adare \textit{et al.} [PHENIX], Phys. Rev. C \textbf{93} (2016) no.5, 051902 [arXiv:1412.1038 [nucl-ex]].
\bibitem{ALICE:2018lao} S.~Acharya \textit{et al.} [ALICE], Phys. Lett. B \textbf{784} (2018), 82-95 [arXiv:1805.01832 [nucl-ex]].

\bibitem{ALICE:2011ab} K.~Aamodt \textit{et al.} [ALICE], Phys. Rev. Lett. \textbf{107} (2011), 032301 [arXiv:1105.3865 [nucl-ex]].
\bibitem{Zhu:2016qiv} X.~Zhu, Adv. High Energy Phys. \textbf{2016} (2016), 4236492 [arXiv:1607.04003 [nucl-th]].
\bibitem{Pandey:2021ofb} S.~Pandey and B.~K.~Singh, J. Phys. G \textbf{49} (2022) no.9, 095001 [arXiv:2107.01880 [hep-ph]].
\bibitem{Schenke:2019ruo} B.~Schenke, C.~Shen and P.~Tribedy,
Phys. Rev. C \textbf{99} (2019) no.4, 044908 [arXiv:1901.04378 [nucl-th]].
\bibitem{Voloshin:2002wa} S.~A.~Voloshin, Nucl. Phys. A \textbf{715} (2003), 379-388 [arXiv:nucl-ex/0210014 [nucl-ex]].

\bibitem{Molnar:2003ff} D.~Molnar and S.~A.~Voloshin, Phys. Rev. Lett. \textbf{91} (2003), 092301 [arXiv:nucl-th/0302014 [nucl-th]].
\bibitem{Singh:2017fgm} A.~Singh, P.~K.~Srivastava, O.~S.~K.~Chaturvedi, S.~Ahmad and B.~K.~Singh, Eur. Phys. J. C \textbf{78} (2018) no.5, 419 [arXiv:1707.07552 [nucl-th]].
\bibitem{Schenke:2020mbo} B.~Schenke, C.~Shen and P.~Tribedy, Phys. Rev. C \textbf{102} (2020) no.4, 044905 [arXiv:2005.14682 [nucl-th]].
\bibitem{Zhu:2015dfa} X.~Zhu, F.~Meng, H.~Song and Y.~X.~Liu, Phys. Rev. C \textbf{91} (2015) no.3, 034904 [arXiv:1501.03286 [nucl-th]].
\bibitem{Bass:1998ca} S.~A.~Bass, M.~Belkacem, M.~Bleicher, M.~Brandstetter, L.~Bravina, C.~Ernst, L.~Gerland, M.~Hofmann, S.~Hofmann and J.~Konopka, \textit{et al.} Prog. Part. Nucl. Phys. \textbf{41} (1998), 255-369 [arXiv:nucl-th/9803035 [nucl-th]].
\bibitem{Qiu:2011hf} Z.~Qiu, C.~Shen and U.~Heinz, Phys. Lett. B \textbf{707} (2012), 151-155 [arXiv:1110.3033 [nucl-th]].
\bibitem{Zabrodin:2016wmo}  E.~E.~Zabrodin, L.~V.~Bravina, B.~H.~Brusheim Johansson, J.~Crkovska, G.~K.~Eyyubova, V.~L.~Korotkikh, I.~P.~Lokhtin, L.~V.~Malinina, S.~V.~Petrushanko and A.~M.~Snigirev, J. Phys. Conf. Ser. \textbf{668} (2016) no.1, 012099. 
\bibitem{PhysRevC.103.014903} S. Pandey, S. K. Tiwari, and B. K. Singh, Phys. Rev. C \textbf{103} (2021), 014903.
\bibitem{Bravina:2020sbz} L.~V.~Bravina, G.~K.~Eyyubova, V.~L.~Korotkikh, I.~P.~Lokhtin, S.~V.~Petrushanko, A.~M.~Snigirev and E.~E.~Zabrodin, Phys. Rev. C \textbf{103} (2021) no.3, 034905 [arXiv:2012.05139 [nucl-th]].
\bibitem{Crkovska:2016flo} J. Crkovska et al., Phys. Rev. C \textbf{95} (2017), 014910; [arXiv:1603.09621[hep-ph]].
\bibitem{CMS:2013wjq} S.~Chatrchyan \textit{et al.} [CMS], Phys. Rev. C \textbf{89} (2014) no.4, 044906 [arXiv:1310.8651 [nucl-ex]]. 
\bibitem{Lokhtin:2005px} I.~P.~Lokhtin and A.~M.~Snigirev, Eur. Phys. J. C \textbf{45} (2006), 211-217 [arXiv:hep-ph/0506189 [hep-ph]].
\bibitem{Lokhtin:2009hs} I.~P.~Lokhtin, L.~V.~Malinina, S.~V.~Petrushanko, A.~M.~Snigirev, I.~Arsene and K.~Tywoniuk, Nonlin. Phenom. Complex Syst. \textbf{12} (2009), 348-355 [arXiv:0910.5129 [hep-ph]].
\bibitem{Lokhtin:2010zz} I.~P.~Lokhtin, L.~V.~Malinina, S.~V.~Petrushanko and A.~M.~Snigirev, Phys. Atom. Nucl. \textbf{73} (2010), 2139-2147.
\bibitem{Lokhtin:2008xi} I.~P.~Lokhtin, L.~V.~Malinina, S.~V.~Petrushanko, A.~M.~Snigirev, I.~Arsene and K.~Tywoniuk, Comput. Phys. Commun. \textbf{180} (2009), 779-799 [arXiv:0809.2708 [hep-ph]].
\bibitem{Amelin:2006qe} N.~S.~Amelin, R.~Lednicky, T.~A.~Pocheptsov, I.~P.~Lokhtin, L.~V.~Malinina, A.~M.~Snigirev, I.~A.~Karpenko and Y.~M.~Sinyukov, Phys. Rev. C \textbf{74} (2006), 064901 [arXiv:nucl-th/0608057 [nucl-th]].
\bibitem{Amelin:2007ic} N.~S.~Amelin, R.~Lednicky, I.~P.~Lokhtin, L.~V.~Malinina, A.~M.~Snigirev, I.~A.~Karpenko, Y.~M.~Sinyukov, I.~Arsene and L.~Bravina, Phys. Rev. C \textbf{77} (2008), 014903 [arXiv:0711.0835 [hep-ph]].
\bibitem{Torrieri:2004zz} G.~Torrieri, S.~Steinke, W.~Broniowski, W.~Florkowski, J.~Letessier and J.~Rafelski,
Comput. Phys. Commun. \textbf{167} (2005), 229-251 [arXiv:nucl-th/0404083 [nucl-th]].
\bibitem{Sjostrand:2006za} T.~Sjostrand, S.~Mrenna and P.~Z.~Skands, JHEP \textbf{05} (2006), 026 [arXiv:hep-ph/0603175 [hep-ph]].
\bibitem{PHENIX:2011yyh}  A.~Adare \textit{et al.} [PHENIX], Phys. Rev. Lett. \textbf{107} (2011), 252301 [arXiv:1105.3928 [nucl-ex]].
\bibitem{ATLAS:2012at} G.~Aad \textit{et al.} [ATLAS], Phys. Rev. C \textbf{86} (2012), 014907 [arXiv:1203.3087 [hep-ex]].
\bibitem{Bravina:2016bei}  L.~Bravina, B.~H.~Brusheim Johansson, J.~Crkovsk\'a, G.~Eyyubova, V.~Korotkikh, I.~Lokhtin, L.~Malinina, E.~Nazarova, S.~Petrushanko and A.~Snigirev, \textit{et al.}
J. Phys. Conf. Ser. \textbf{736} (2016), 012024 [arXiv:1606.03250 [hep-ph]].
\bibitem{ATLAS:2018ezv}  M.~Aaboud \textit{et al.} [ATLAS], Eur. Phys. J. C \textbf{78} (2018) no.12, 997 [arXiv:1808.03951 [nucl-ex]].

\bibitem{PHENIX:2012swz} A.~Adare \textit{et al.} [PHENIX],
Phys. Rev. C \textbf{85} (2012), 064914 [arXiv:1203.2644 [nucl-ex]].


\end{thebibliography}
\end{document}